\def\year{2022}\relax
\documentclass[letterpaper]{article} 
\usepackage{aaai22}  
\usepackage{times}  
\usepackage{helvet}  
\usepackage{courier}  
\usepackage[hyphens]{url}  
\usepackage{graphicx} 
\def\UrlFont{\rm}  
\usepackage{natbib}  
\usepackage{caption} 
\DeclareCaptionStyle{ruled}{labelfont=normalfont,labelsep=colon,strut=off} 
\usepackage{algorithm}
\usepackage{algorithmic}

\usepackage{newfloat}
\usepackage{listings}
\floatstyle{ruled}
\newfloat{listing}{tb}{lst}{}
\floatname{listing}{Listing}
\usepackage{ctable}
\usepackage{mathtools}

\title{ExBrainable: An Open-Source GUI for CNN-based \\ EEG Decoding and Model Interpretation}
\author{
    Ya-Lin Huang,\textsuperscript{\rm 1} 
    Chia-Ying Hsieh,\textsuperscript{\rm 1} 
    Jian-Xue Huang,\textsuperscript{\rm 1} 
    Chun-Shu Wei\textsuperscript{\rm 1}
}

\affiliations{
    \textsuperscript{\rm 1}National Yang Ming Chiao Tung University, Hsinchu, Taiwan\\
    yalinhuang.bt06@nycu.edu.tw, irishsieh0720.cs07@nycu.edu.tw, abcdefg101074@gmail.com, wei@nycu.edu.tw
}

\usepackage{bibentry}

\begin{document}
\maketitle
\begin{abstract}
We have developed a graphic user interface (GUI), ExBrainable, dedicated to convolutional neural networks (CNN) model training and visualization in electroencephalography (EEG) decoding. Available functions include model training, evaluation, and parameter visualization in terms of temporal and spatial representations. We demonstrate these functions using a well-studied public dataset of motor-imagery EEG and compare the results with existing knowledge of neuroscience. The primary objective of ExBrainable is to provide a fast, simplified, and user-friendly solution of EEG decoding for investigators across disciplines to leverage cutting-edge methods in brain/neuroscience research.
\end{abstract}

\section{Introduction}
Electroencephalography (EEG) has been a neuromonitoring modality widely used for mind, brain and neuroscience research, and to facilitate brain-computer interfacing (BCI). The advantages of EEG in terms of practical application development include non-invasiveness, affordability, and high temporal resolution. These characteristics combined allow EEG devices a feasible solution in real-world scenario for daily use \cite{koike2013near,mehta2013neuroergonomics,sejnowski2014putting}. In general, EEG signals are measured by a number of electrodes placed on the scalp. Each electrode senses subtle fluctuation of electrical field driven by the neuroelectrophysiological activity of a population of neurons in the local cortical brain \cite{cohen2017does}. The EEG signals recorded by multiple electrodes form multi-channel time series that carry information in spatial (across brain regions) and temporal (across time) domain. Yet the relatively low signal-to-noise ratio remains an inevitable challenge in decoding brain activity through EEG signals \cite{johnson2006signal,bast2006noninvasive}. 

The fast growth of deep learning methods have largely eliminated the need for handcrafted feature extraction and achieved state-of-the-art performance in numerous fields \cite{lecun2015deep}. Convolutional neural networks (CNN) particularly are able to mimic human visual perception and can perform a variety of feature extractions in problems of computer vision and automatic speech recognition \cite{lecun1995convolutional,krizhevsky2012imagenet,albawi2017understanding}. Lately, the use of CNN models extends to other domain including EEG data decoding and analysis. Customized CNN models for EEG decoding have outperformed conventional approaches and significantly reduced laborious manual feature extraction \cite{lawhern2018eegnet,HBM:HBM23730,wei2019spatial, krizhevsky2012imagenet}. These CNN models are able to handle minimally processed multi-channel EEG signals, and thus enable end-to-end EEG decoding for translating brain activities to meaningful information such as the commands for cursor control \cite{nicolas2012brain}. CNN-based EEG decoding has drawn promising outcome in recognizing various types of EEG data including rhythmic brain oscillation during motor imagery \cite{HBM:HBM23730,wei2019spatial} and event-related potentials \cite{lawhern2018eegnet,waytowich2018compact}. 

While CNN models enhance the performance of EEG decoding, it has been of great interest that what a CNN model can learn from EEG data. Without proper transparency and interpretation, the CNN models could turn out to be "black boxes", hindering the understanding of underlying scaffold of these advanced computational approaches. Efforts have been made to visualize parameters and latent features within the CNN models fitted to EEG data. For instances, the kernels preforming convolution along the temporal dimension operate as temporal filters and extract EEG features in time. The temporal convolutional kernels can be visualized by waveform observation and spectral analysis to characterize in time and frequency domain \cite{lawhern2018eegnet}. On the other hand, the convolutional kernels operate in spatial domain are regarded as spatial filters that combine signals across channel to enhance signals or reduce noise and artifacts. The topography of scalp is a straightforward tool to convey the spatial distribution of kernel weights \cite{HBM:HBM23730,lawhern2018eegnet}. These visualization techniques adequately transform complex CNN models into "glass boxes", providing understandable and explainable representations that are compatible to conventional insight of neuroscience. In summary, current development of CNN-based EEG decoding have major strengths including 1) rapid EEG data analysis and decoding without tedious hand-crafted feature engineering; 2) data-driven modeling of EEG data that tolerates elusive variability due to individual difference and non-stationarity; and 3) Explainable model interpretation for accelerating data analysis and further improvement in model design. Yet there is lack in this field of an open and user-friendly software that integrates cutting-edge techniques for explainable AI-based EEG data modeling, decoding, and interpretation.

We introduce ExBrainable, an open, compact, and easy-to-use tool for EEG investigators to leverage the cutting-edge computational algorithms and functions for CNN-based EEG decoding and model interpretation with almost zero programming prerequisites. The user-friendly graphic interface based on Python allows smooth operation of modeling pipeline and collection of feature interpretation. Constructed based on Python scripting and multiple open libraries, ExBrainable is completely free available for investigators of different disciplines with limited resource and limited programming skill. In this paper, we present ExBrainable by providing background information about existing open software for EEG analysis and decoding, followed by an overview of functions and features of ExBrainable. Next, we demonstrate an evaluation of ExBrainable using well-known public EEG dataset with an in-depth analysis on the experimental results and on the model interpretation. Finally, we summarize the contributions made in this work as well as future work for further improvement and extension.

\section{Related Work}
This section briefs representative brain analysis tools that include functions based on machine learning approaches. 

\subsection{EEGLAB}
In 2004, Delorme and Makeig \cite{delorme2004eeglab} developed EEGLAB, an open-source toolbox running in MATLAB environment (The Mathworks, Inc.) for EEG processing. The features include EEG data preprocessing, cleaning, filtering, and visualization. It provides independent component analysis (ICA) that observes brain activity in a projected component domain, and emphasizes time-frequency representation of brain dynamics. Later on, the addition of dipole fitting functions enables analysis of brain source on the basis of 3-D brain regions. Numerous plug-in functions released by third party developer enhances the flexibility and versatility of EEGLAB. The user-friendly graphic interface makes EEGLAB has become one of the most widely used EEG analysis tool liked by investigators across multiple disciplines. 

\subsection{BCI2000}
BCI2000 \cite{schalk2004bci2000} is a general-purpose platform for BCI research and development. BCI2000 can process brain signals using signal processing methods, output devices and operating protocols. The essential modules of BCI2000 include source, signal processing, user application, and operator. For BCI implementation, BCI2000 has multiple strengths including interchangeability, scalability, real-time capability, support of offline analyses, and practicality. It is able to reduce labor and cost in online operation of real-time BCI systems as well as a variety of psychophysiological experiments, and is free of charge for research or educational purposes.

\subsection{Brainstorm}
Brainstorm \cite{tadel2011brainstorm} is an open-source application for EEG/MEG data processing and visualization. Brainstorm is developed using Matlab scripts and distributed under the terms of the General Public License, with interface written in Java/Swing embedded in Matlab scripts. As a fully portable, cross-platform application, Brainstorm provides graphic user interface for investigators to access all functions without programming. Brainstorm features extensive options of preprocessing method, interactive interface for data visualization, and individualized anatomy importation of MRI information.

\subsection{MNE-Python}
MNE-Python \cite{gramfort2013meg} is an open-source software implemented as a Python package that provides multiple methods for EEG/MEG data processing. The major features includes data preprocessing techniques, source localization, statistical analysis, and estimation of cross-brain-region functional connectivity. The algorithms and functions are designed to form a consistent and well-documented interface for painless construction of EEG/MEG processing pipeline with integration of scientific computation and visualization. Although part of the functions allows operations in graphic interfaces, MNE-Python mainly requires Python programming in order to fully utilize its functions and features.

\subsection{Braindecode}
Braindecode \cite{HBM:HBM23730} is a deep learning toolbox recently developed for end-to-end EEG decoding using processed multi-channel time series. Unlike the previous tools that focus mainly on the preprocessing pipeline, Braindecode performs deep-learning modeling of EEG data targeting the needs of EEG decoding with cutting-edge schemes. The primary features of Braindecode include EEG data loading, preprocessing, segmentation, training/validation set splitting, model training, and presentation of decoding results. The current version of Braindecode incorporates many existing functions directly from MNE-Python, and thus programming skill is required for a painless operation.

\section{Design Approaches}

\begin{figure*}[t!]
\centering
\includegraphics[width=17cm]{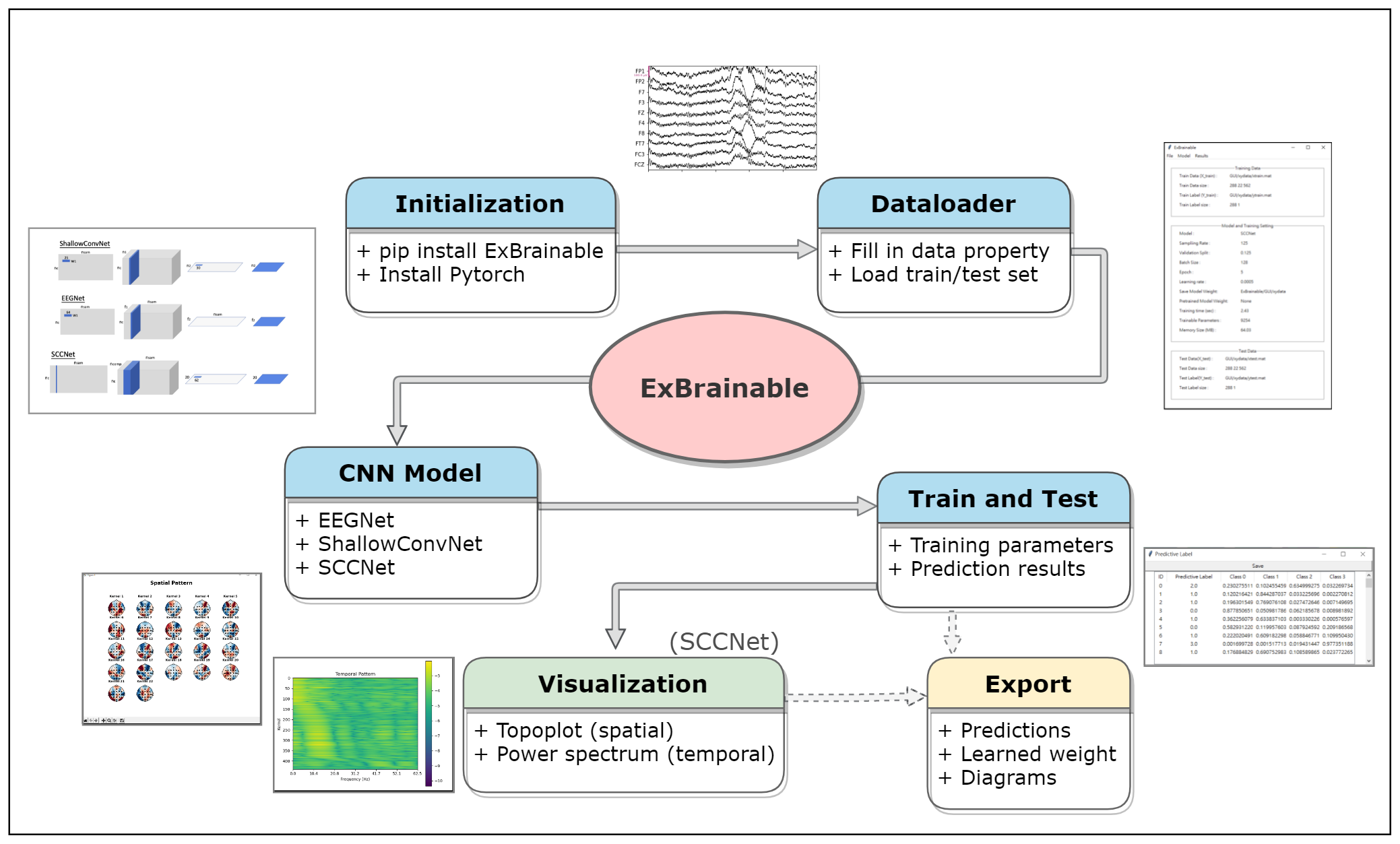}
\caption{Schematic Flow of ExBrainable}
\label{flowchart}
\end{figure*}

The major considerations in the design of ExBrainable include 1) fast implementation and application of CNN-based EEG decoding; 2) Explicit interpretation of model parameters; and 3) an easy-to-use GUI without any programming operations. We adopted several useful open tools and libraries based on Python in the development of ExBrainable.

\subsection{CNN Model Implementation and Training}
ExBrainable provides built-in Python scripts of CNN models with functions of model fitting and evaluation using PyTorch \cite{paszke2019pytorch}, an open source machine learning framework. PyTorch allows developers to construct end-to-end workflow in neural network models and supports accelerated parallel computing platform such as Compute Unified Device Architecture (CUDA). In addition, developers can optimize large-scale networks using computational graph at each execution point \cite{ketkar2017introduction}.

\subsection{Visualization of Model Parameter}
MNE-Python \cite{gramfort2013meg} provides a broad selections of functions for EEG processing, analysis, and visualization. In the block of model interpretation in ExBrainable, the visualization of spatial convolutional kernels uses \texttt{mne.viz.plot{\_}topomap} to generate color topographic scalp map as a intuitive presentation of weight distribution across EEG channels. The visualization of temporal convolutional kernels uses the \texttt{fft} functions in the well-known SciPy package \cite{virtanen2020scipy}, a open source Python library for scientific and technical computing.

\subsection{Graphic Interface}
The GUI of ExBrainable is developed with Tkinter \cite{lundh1999tkinter}, a Python interface to the Tcl/Tk GUI toolkit for building  graphic interface of applications. Tkinter provides a variety of widgets for general purposes of GUI development. Using TKinter, developers can easily integrate other Python-based functions and scripts and facilitate backend operations.

\section{Features and Functions}
In this section, we specify the main features and functions of ExBrainable by blocks in the data modeling pipeline. The main dashboard display the status of the modeling pipeline with information about data loading, model training setting, and model status. The menu includes the options of operations to be performed in the pipeline as shown in Figure \ref{flowchart}.

\begin{figure}[t!]
\centering
\includegraphics[width=7cm]{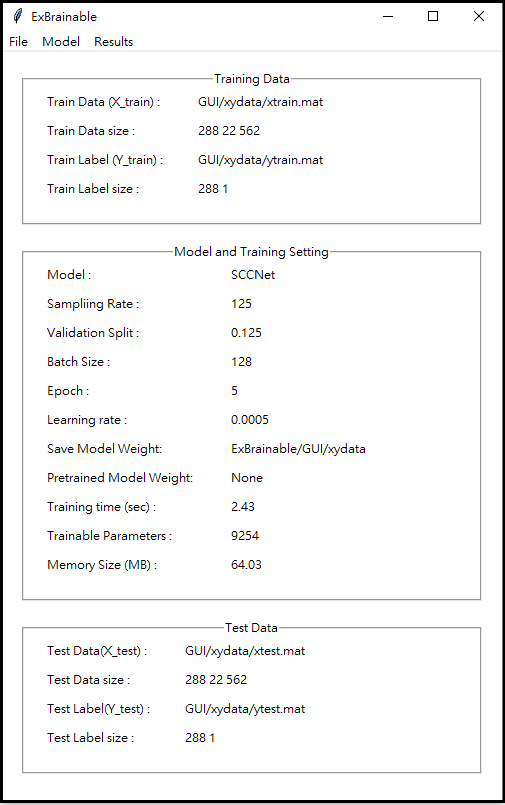}

\caption{Main dashboard of the ExBrainable GUI.}
\label{GUI}
\end{figure}

\subsection{Data Loader}
It is necessary to have training data for CNN-based EEG decoding. ExBrainable loads EEG data and labels for model training, model prediction, and performance evaluation. ExBrainable takes EEG data undergone preprocessing with no restriction. Users can utilize the above-mentioned EEG preprocessing tools such as EEGLAB, BCI2000, Brainstorm, or MNE-Python to perform data cleaning and segmentation. If the data format fits properly, ExBrainable will activate the functions of model training (with training set available), prediction (with test data available) and performance evaluation (with test label avaialble). 

\subsection{Built-in Model Base}

ExBrainable has a built-in model base providing representative convolutional neural network architectures available for users to load and train the selected model. The models currently on shelf are ShallowConvNet \cite{HBM:HBM23730}, EEGNet \cite{lawhern2018eegnet}, and SCCNet \cite{wei2019spatial} implemented in PyTorch scripts. The user can also use other models by incorporating codes compatible to the script of the built-in model base. The detail of model architectures is provided as in the supplementary material. 

\subsubsection{ShallowConvNet}
ShallowConvNet was proposed simultaneously with DeepConvNet for decoding motor-imagery EEG data in same published work \cite{HBM:HBM23730}. Both CNN models were inspired by filter-bank common spatial pattern \cite{ang2008filter}, a conventional spatial filtering approach, and have performed on par on the decoding of motor-imagery EEG. Therefore, ShallowConvNet has been more popular than DeepConvNet because of its smaller model size. ShallowConvNet takes multi-channel EEG signals as inputs and performs temporal convolution in its first layer, followed by a second convolutional layer across space (channels).

\subsubsection{EEGNet}
EEGNet has a compact CNN architecture customized for extracting intrinsic features of EEG  \cite{lawhern2018eegnet}. EEGNet can process multi-channel EEG time series as input, using a temporal convolutional layer that acts as a series of spectral filters and other temporal operations, followed by a depthwise spatial convolutional layer that serves as spatial filters for signal enhancement and dimensionality reduction. The use of separable convolutions elevates the efficiency of EEGNet while maintaining a small model size. EEGNet performs nicely on various types of EEG data including event-related potentials and rhythmic activities \cite{lawhern2018eegnet,waytowich2018compact}. 

\subsubsection{SCCNet}
Proposed in 2019, SCCNet \cite{wei2019spatial} first process the multi-channel EEG data using a spatial convolutional kernels to characterize data in the spatial domain. The spatial convolutional operations are able to reduce noise and to enhance signals in EEG data as many spatial filtering techniques do (e.g. montage, referencing, component analysis). Following by spatial convolution, a spatial-temporal convolutional layer performs temporal filtering and cross-component combination of the latent features generated by the first layer. Analogues to ShallowConvNet and EEGNet, SCCNet features small model size and is able to fit relatively small EEG datasets.

\subsection{Model Training}

Once a CNN model has been selected, the model training process can start after filling in parameters such as batch size, number of training epochs, learning rate, and the saving path of model weights. Details of model training are available in the online documents of PyTorch \cite{paszke2019pytorch}.




\subsection{Model Prediction and Performance}

The prediction results are shown as a table with the probability (logit output) of each class and the predicted label determined by the class with the maximal probability. The output label and class probability can be saved as a .csv file with one click. If the ground truth labels are available, the model performance will be evaluated by classification metrics including accuracy and Cohen's kappa. In addition, the confusion matrix will be available to exhibit the detail of prediction.

\begin{figure}[t!]
\centering
\includegraphics[width=7cm]{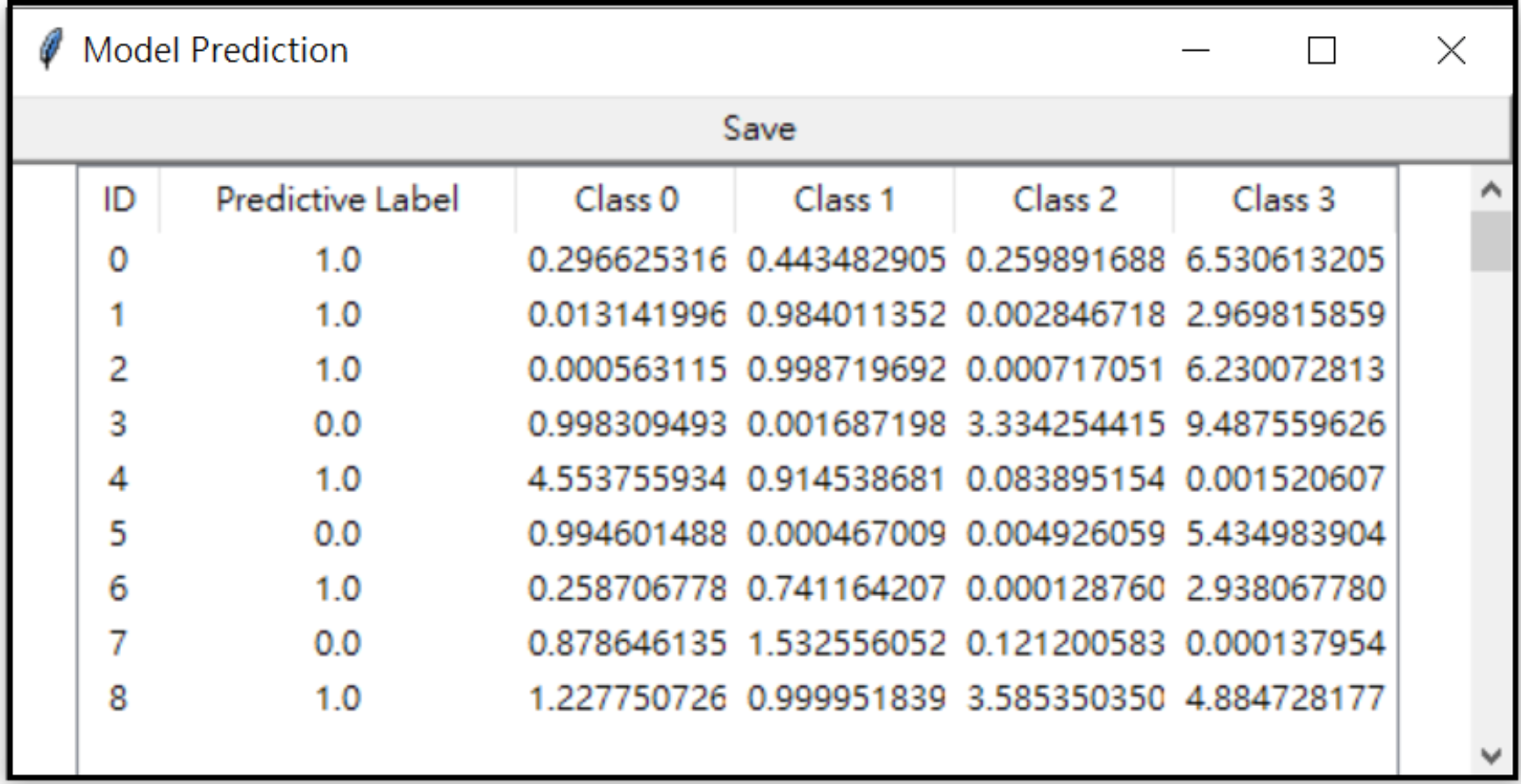}

\caption{The output of model prediction using ExBrainable. 'ID' denotes the index of test data sample. 'Predicted Label' is the class label with the maximal. }
\label{GUI_label}
\end{figure}

\begin{figure}[h]
\centering
\includegraphics[width=6cm]{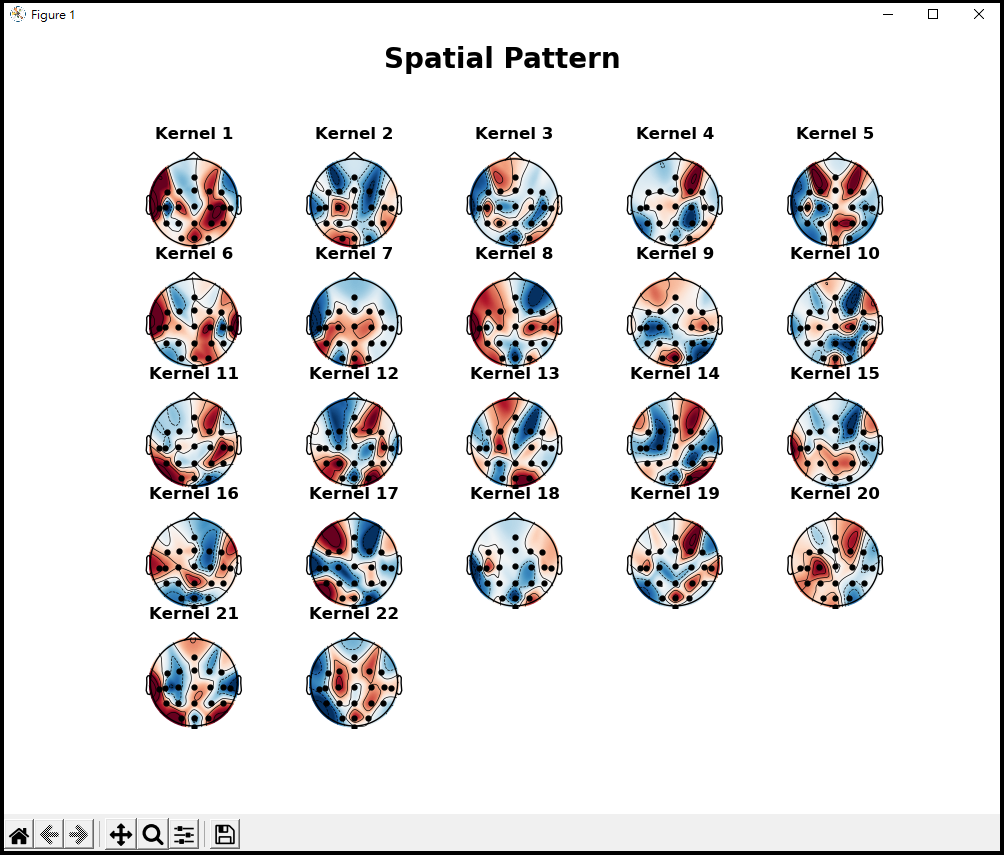}
\caption{Spatial patterns learned by the model.}
\label{spatial_learned}
\end{figure}
\begin{figure}[h]
\centering
\includegraphics[width=6cm]{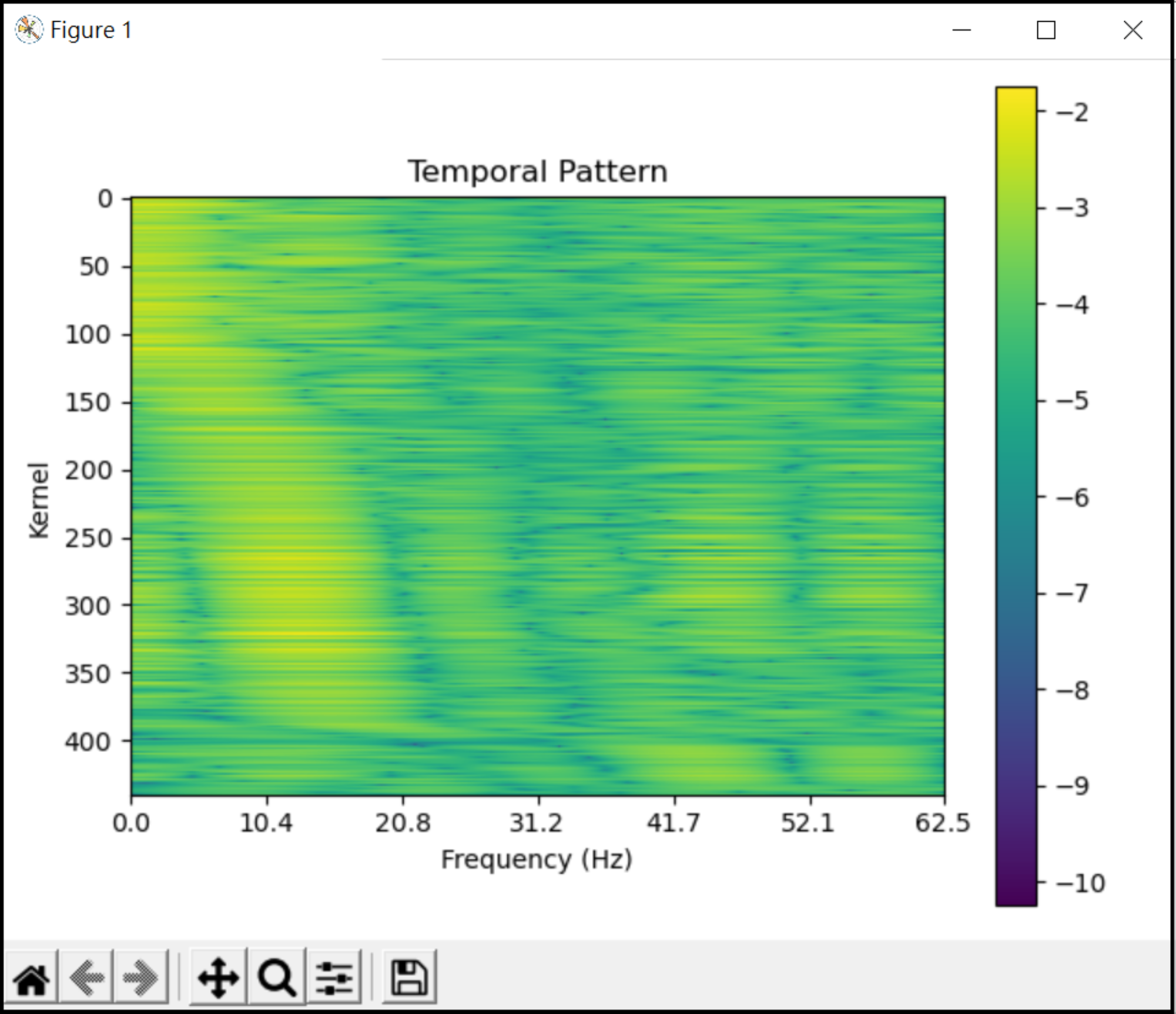}
\caption{Temporal patterns learned by the model.}
\label{spatial_learned}
\end{figure}

\subsection{Model Interpretation for SCCNet}
The current version of ExBrainable provides visualization functions for pre-trained SCCNet models. The SCCNet architecture starts from the first layer of spatial convolutional kernels. Each spatial convolutional kernel represents a linear combination of channels for the EEG data to project. As the weights in a spatial convolutional kernel are corresponding to the channels on the scalp, scalp topography clearly illustrates the spatial distribution of the weights. The second layer incorporates temporal convolution to extract the temporal features on the latent features. One of the most important characteristics to observe is the frequency response of the temporal convolutional kernels. To provide a complete view of spectral properties within the temporal convolutional kernels, the frequency responses of temporal kernels are sorted and presented as an image where the bright color marks the high spectral response. The kernel weights of spatial convolutional kernels and temporal convolutional kernels are available through one click and can be saved as picture files if needed.

\section{Evaluation}
ExBrainable features user-friendly GUI, fast implementation of a CNN-based EEG decoding flow, and intuitive visualization of EEG data insights. We herein performed an evaluation for ExBrainable using the BCI competition IV-2a (BCIC-IV-2a) dataset \cite{brunner2008bci}, a well-studied public dataset of motor-imagery EEG. A series of simulations were conducted to demonstrate the functions and assess the decoding performances of the built-in model base in ExBrainable. 

\subsection{Data}
The BCIC-IV-2a dataset \cite{brunner2008bci} consists of motor-imagery (MI) EEG data collected in a motor-imagery experiment of nine subjects, where subjects were asked to imagine moving different body parts (right hand, left hand, feet, and tongue). One experimental session includes 72 trials of each class of motor imagery, and 22 channels of EEG time series were measured simultaneously. The dataset consists of EEG recordings from nine participants. Each participant performed two sessions on different days. The montage of EEG recording is available in the supplementary material.


\subsection{Experiments}
Based on the practical scenario of BCI usage, we adopted four training schemes that serves for different conditions of practical BCI usage: 1) Individual training, 2) Subject-independent (SI) training, 3) Subject-dependent (SD) training, and 4) Subject independent training plus fine-tuning (SI+FT) {\cite{wei2019spatial}}. The four training schemes are described as following:

\begin{itemize}
\item
Individual: The individual training scheme refers to dividing training and test data within a single subject. In the BCIC-IV-2a dataset, the first session of a subject served as the training data, and the model was evaluated on the second session of the same subject. This training scheme is applied when a BCI user performs a training session before executing the BCI.
\end{itemize}

\begin{itemize}
\item
SI: The SI training scheme is a way to train a model without any information from a target user. This scheme does not require any data from the test subject, but use all of the data from other subjects. For BCI usage, the SI training scheme is able to generate a pre-trained model available for a new user without any training/calibration.
\end{itemize}

\begin{itemize}
\item
SD: The SD training scheme incorporates all of the data from other subjects concatenated on the first session of the test subject. This scheme is applicable when existing data are available and the target user has completed a training session before usage.
\end{itemize}

\begin{itemize}
\item
SI+FT: The SI+FT training scheme first uses the SI training scheme to train the model, and then applies fine-tuning process on the pre-trained model with first session of the test subject. This scheme utilize exactly the same training data as in the SD training scheme, but it involves two phases of model training rather than a single phases in the SD training scheme. Therefore, the usage of the SI+FT training scheme requires the same condition as for the SD training scheme where a training session is needed for the new user, while a pre-trained model based on all of the data from other subjects should be prepared in advance.
\end{itemize}

Using the dataset BCIC-IV-2a, we trained the three models in the built-in model base (EEGNet, ShallowConvNet, and SCCNet) and evaluate their performances based on the four training schemes. The computational experiments were completed using a NVIDIA GeForce RTX 2080 Ti 11G GPU. Parameters in the training setting were chosen empirically as described in the supplementary material. The model fitting procedure used a batch size of 128, a learning rate of $5\times10^{-4}$, a validation split of 0.125 for 500 episodes.
\begin{table}[b!]
\centering
\resizebox{\linewidth}{!}{ 
\renewcommand{\arraystretch}{1.2}
{\begin{tabular}{cccc}
\specialrule{.1em}{.1em}{.1em}
    &  EEGNet & ShallowConvNet & SCCNet \\
\hline
    Individual & 0.63±0.15 & 0.72±0.12 & 0.78±0.13\\
    SD  & 0.49±0.17 & 0.67±0.13 & 0.68±0.13\\
    SI  & 0.43±0.15 & 0.50±0.15 & 0.49±0.15\\
    SI+FT & 0.63±0.13 & 0.75±0.12 & 0.79±0.12\\ 
\specialrule{.1em}{.1em}{.1em}
\end{tabular}}}
\caption{EEG decoding accuracy of the CNN models under individual, subject-dependent (SD), subject-independent (SI), and SI plus fine-tuning (SI+FT) training schemes.}
\label{table_accuracy}
\end{table}

\begin{table}[b!]
\centering
\renewcommand{\arraystretch}{1.2}
\resizebox{\linewidth}{!}{ 
\setlength{\tabcolsep}{3.5mm}{ 
{\begin{tabular}{lllll}
\specialrule{.1em}{.1em}{.1em}
     & Individual & SD & SI & SI+FT\\
\hline
    (a) {$>$} (b) &  &  &  & \\
    (a) {$>$} (c) &  &  &  & \\ 
    (b) {$>$} (a) & * & ** & ** & **\\
    (b) {$>$} (c) &  &  &  & \\
    (c) {$>$} (b) & ** &  &  & **\\
    (c) {$>$} (a) & ** & ** &  & **\\
\specialrule{.1em}{.1em}{.1em}
\multicolumn{2}{l}{
*$p < 0.05$, **$p < 0.01$}
\end{tabular}}}}
\caption{Comparison of EEG decoding accuracy between models ((a) EEGNet; (b) ShallowConvNet; (c) SCCNet) using Wilcoxon Signed rank test.}
\label{table_statistics}
\end{table}

\subsection{EEG Decoding Performance}
This section exhibits the results of model performance gained from the demonstration of ExBrainable based on the experiments described above.
The classification performance of motor-imagery EEG data was evaluated by the averaged accuracy based on 30 evaluation runs for each target subject. Table \ref{table_accuracy} summarizes the accuracy for all three CNN models under different training schemes, and the corresponding statistical analysis for model comparison is available in Table \ref{table_statistics}. Among all three models, SCCNet significantly outperforms both EEGNet and ShallowConvNet under the individual and SI+FT schemes, and outperforms only the EEGNet under the SD scheme. ShallowConvNet yields the best accuracy under the SI scheme, yet the difference to SCCNet is marginal. In the comparison between ShallowConvNet and EEGNet, ShallowConvNet outperforms EEGNet under all schemes. 

Besides the choice of model, the accuracy varies by the training schemes. The individual and SI+FT schemes are with higher accuracy compared to the SD and SI schemes. This discrepancy across training schemes in the decoding accuracy is in line with the previous study \cite{wei2019spatial}. As reported, the training process of the SD and SI schemes may face the issue of individual difference in the motor-imagery EEG data. The heterogeneity in the combined training set across subjects could cause deteriorated training. On the other hand, individual training performed within subject avoids the cross-subject variability. Fine-tuning a pre-trained SI model is a way to further enhance the performance, possibly because it leverages the informative pre-trained weights from the SI model and facilitates transfer learning process to fit the individualized data in an extra training phase \cite{wei2019spatial}. 

Another aspect to consider regarding the choice of model is the time cost for model training. We recorded the training time to display on the dashboard of ExBrainable, as well as the related parameters, the size of trainable parameters and the size memory required for model training. The comparison of training time under different schemes are summarized as in Table \ref{table_training_time}. Overall, ShallowConvNet has the most trainable parameters and memory size, and required the most training time under all schemes. SCCNet, although has an intermediate size of trainable parameters among the three models, utilizes the least memory size and the shortest training time under all schemes. Intuitively, we expected a short training time associates with a small size of trainable parameters, yet the observation suggests that the training time might have an even stronger relationship to the memory size rather than the trainable parameters.

\begin{table}[t!]
\centering
\resizebox{\linewidth}{!}{ 
\renewcommand{\arraystretch}{1.2}
{\begin{tabular}{cccc}
\specialrule{.1em}{.1em}{.1em}
    &  EEGNet & ShallowConvNet & SCCNet \\
\specialrule{.1em}{.1em}{.1em}
    Individual & 0 m 13 s & 0 m 24 s & 0 m 7 s\\
    SD  & 1 m 41 s & 3 m 48 s & 0 m 41 s\\
    SI  & 1 m 29 s & 3 m 19 s & 0 m 39 s\\
    SI+FT & 1 m 51 s & 3 m 54 s & 0 m 41 s\\ 
\hline
    {\begin{tabular}{c}
        {Trainable}\\
        {Parameters}
    \end{tabular}} 
    & 2,548 & 47,644 & 9,254\\
\specialrule{.1em}{.1em}{.1em}
    {\begin{tabular}{c}
        {Memory}\\
        {size(MB)}
    \end{tabular}}
    & 239.69 & 572.63 & 64.03\\
\specialrule{.1em}{.1em}{.1em} 

\end{tabular}}}
\caption{Training time of subject 1 on 4 schemes experiment with EEGNet, ShallowConvNet, SCCNet}
\label{table_training_time}
\end{table}

\begin{figure*}[t!]
\centering
\frame{\includegraphics[width=6cm]{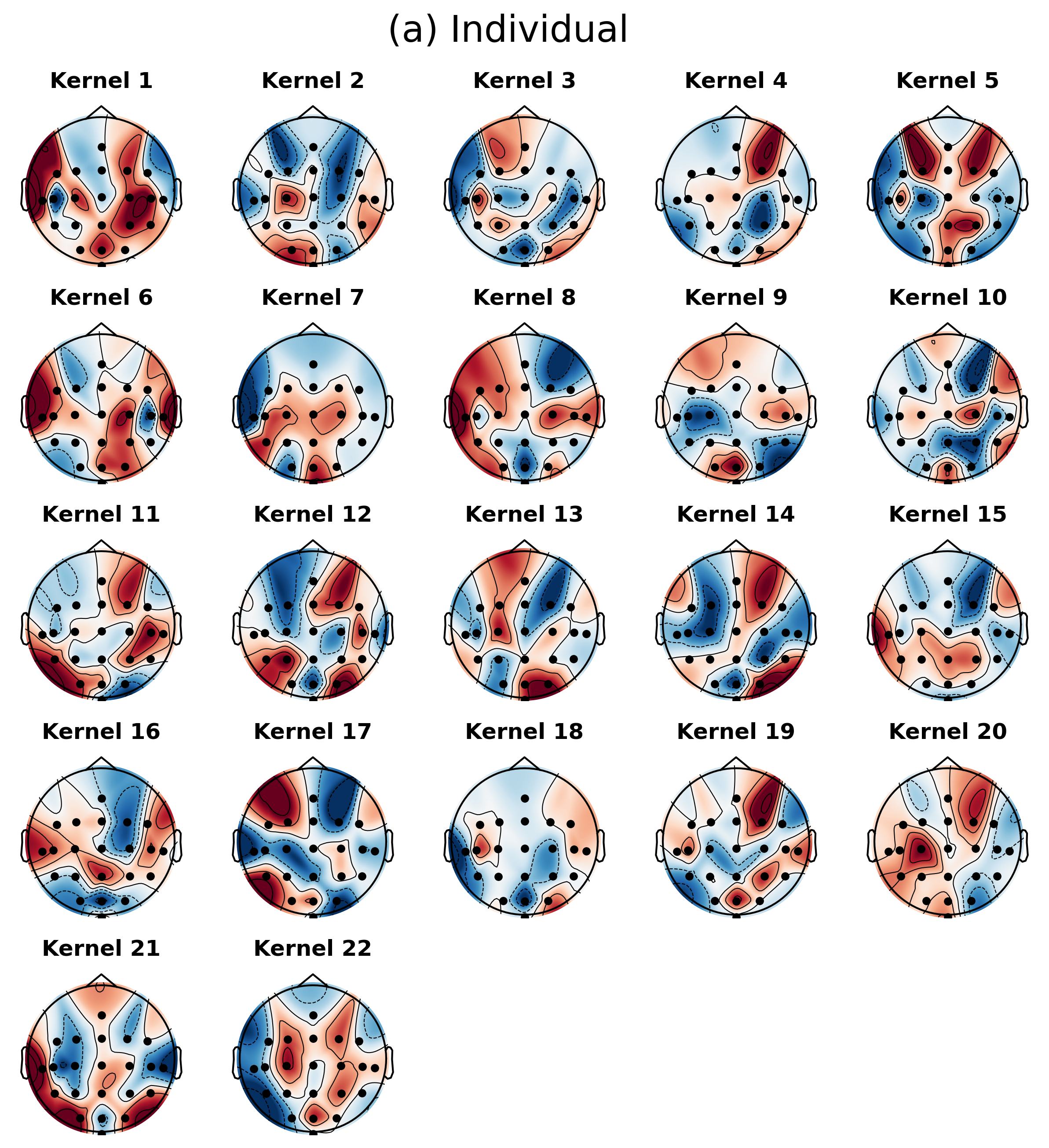}}
\quad
\frame{\includegraphics[width=6cm]{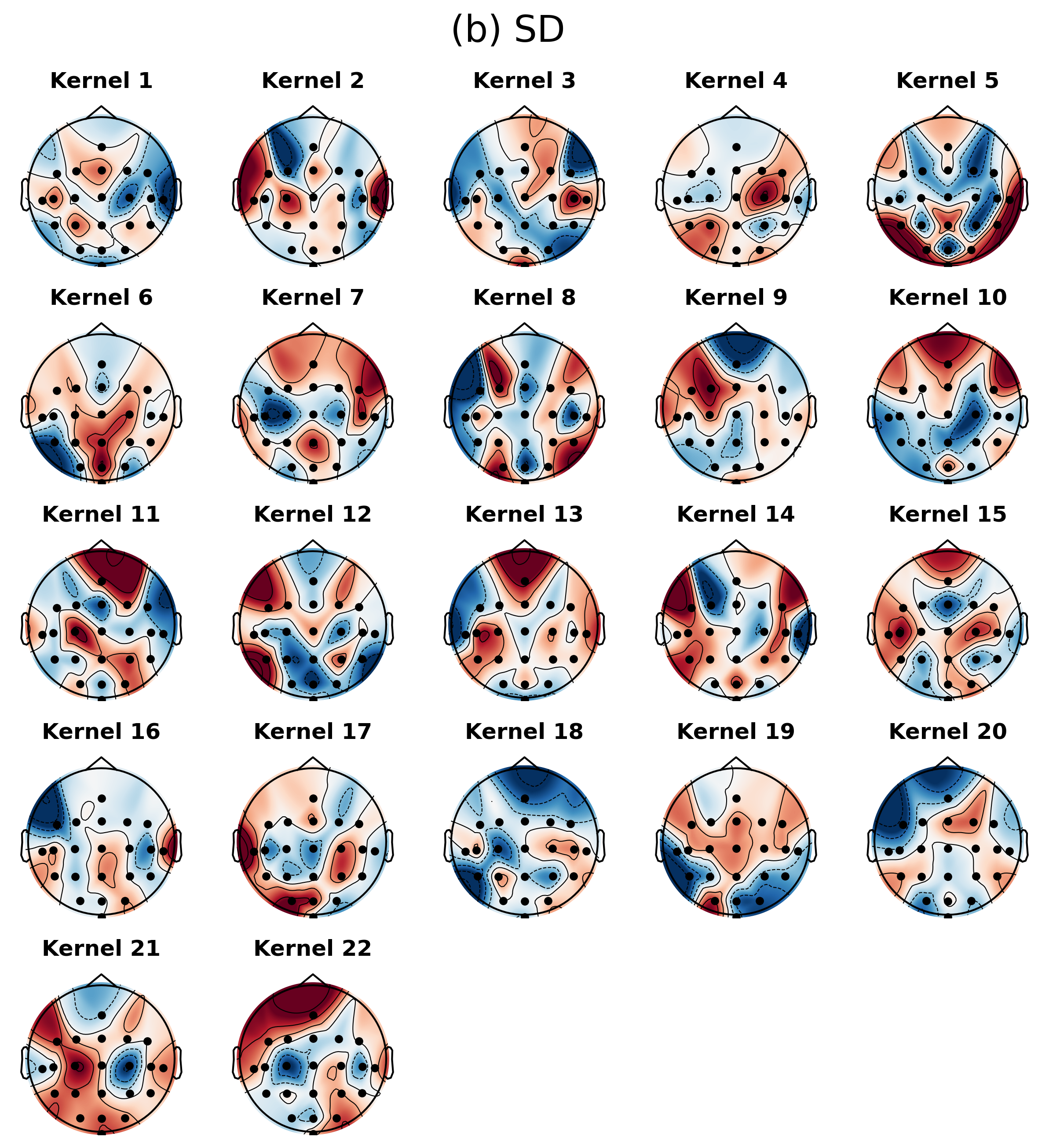}}

\vspace{0.4cm}

\frame{\includegraphics[width=6cm]{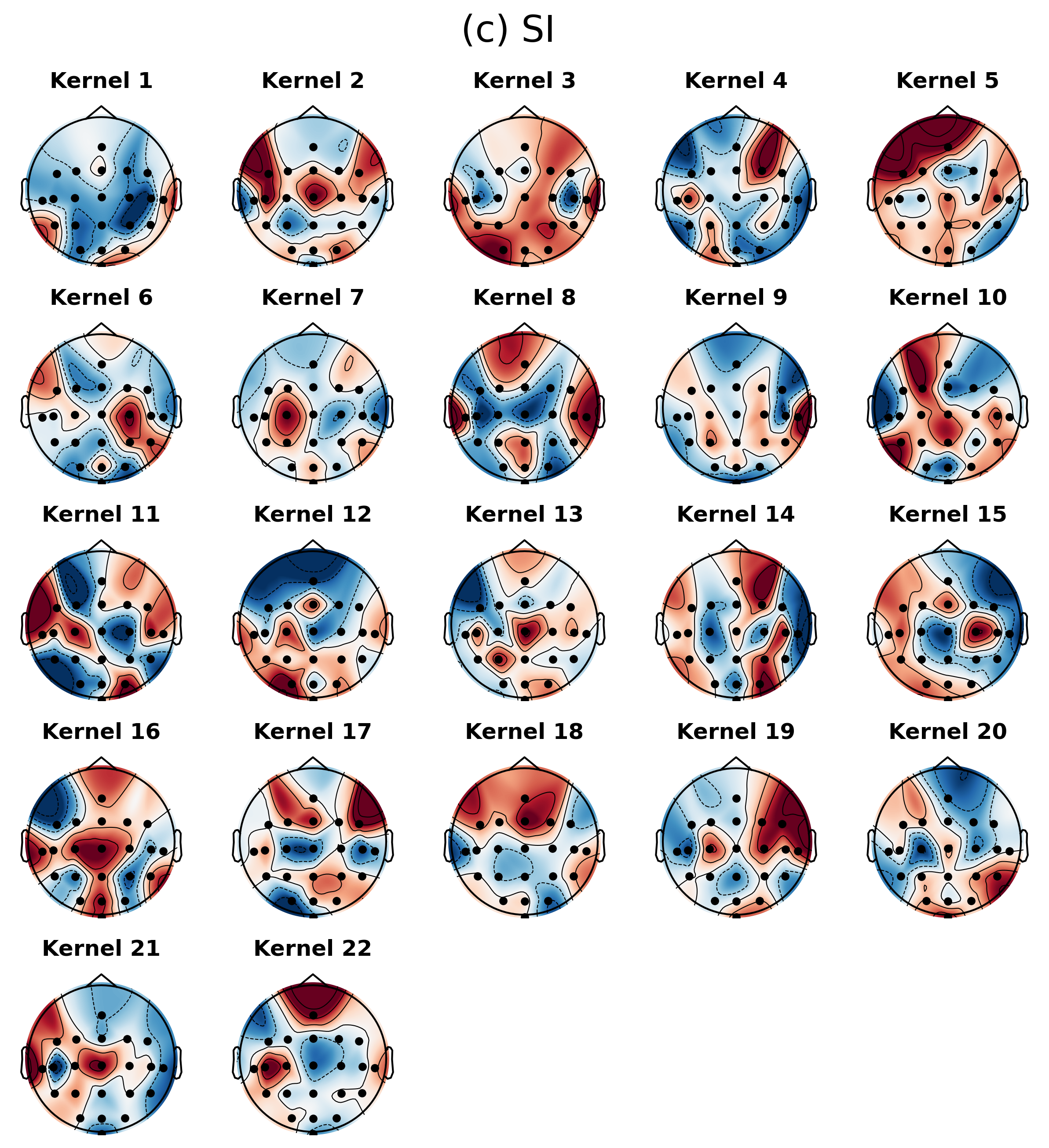}}
\quad
\frame{\includegraphics[width=6cm]{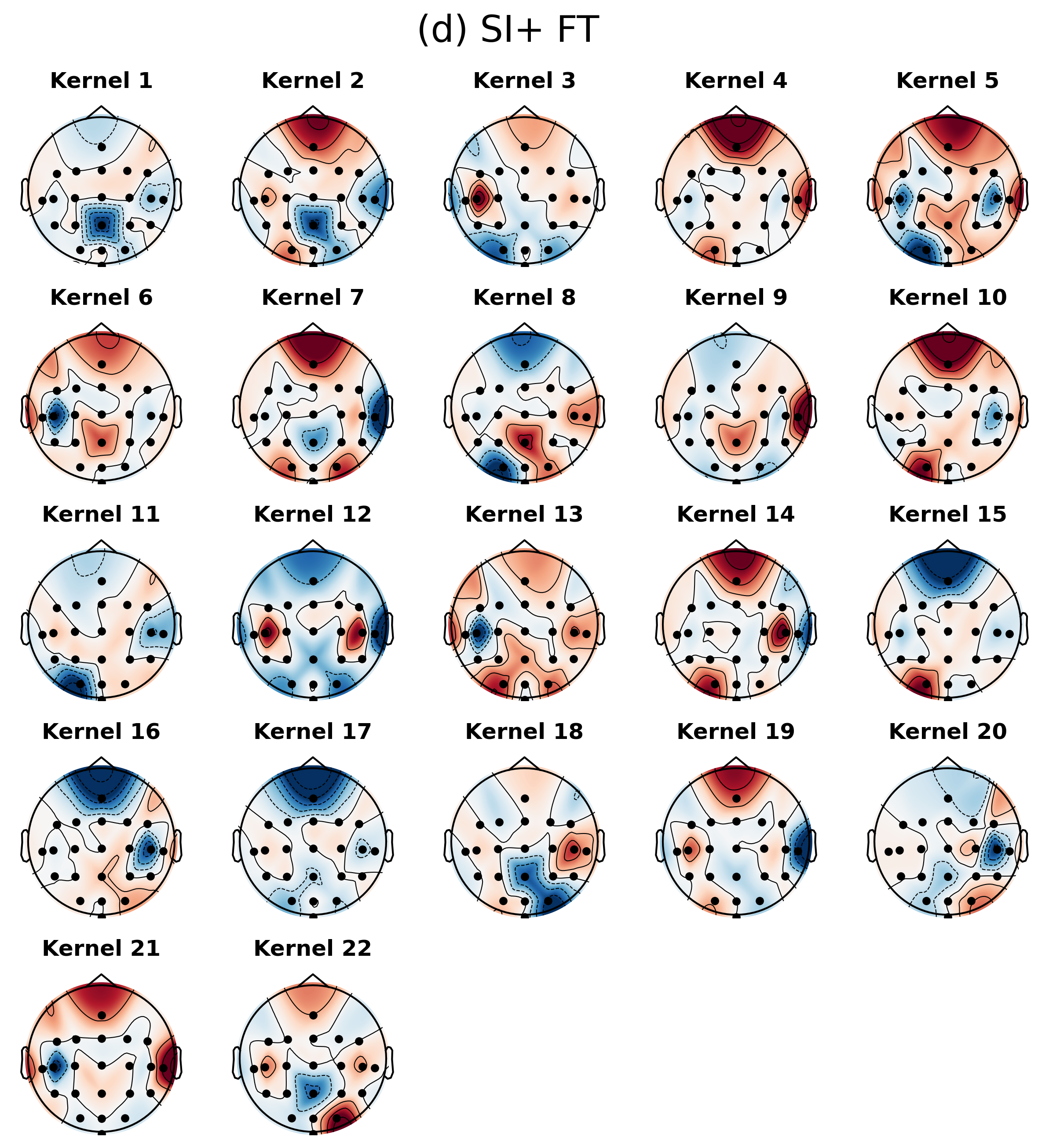}}

\vspace{0.4cm}

\caption{Visualizing learned weights in first convolutional layer of SCCNet: Topographic scalp maps of spatial kernels for subject 3 on 4 training schemes. (a) Individual; (b) Subject Dependent; (c) Subject Independent ; (d) Subject Independent+ Fine-Tuning.}
\label{spatial_patterns}
\end{figure*}

\subsection{Model Interpretation}
We illustrate the two approaches to interpret the parameters learned by SCCNet: (1) Visualizing kernel weights in the (first) spatial convolutional layer, (2) Summarizing the spectral patterns in the temporal convolutional kernels. Visualization of spatial kernels under different training schemes is shown in Figure \ref{spatial_patterns}. Visualization of temporal kernels with in-depth interpretation and discussion is available in the supplementary material.

\section{Conclusion and Future Work}
In this paper, we developed ExBrainable, a CNN model training and model interpretation GUI with functions to visualize learnt kernel weights as topographical maps and power spectrums for EEG-based CNN model interpretation. Furthermore, we found that SCCNet has better performance in both classification accuracy and training time cost comparing to EEGNet and ShallowConvNet, hence SCCNet could be promising to be employed in future real-time BCI system developments.







\section{Video Demonstration and Code}
The brief demo of ExBrainable is on \url{https://www.youtube.com/watch?v=m40z2klbmtg}. If you want to use ExBrainable, please direct to \url{https://github.com/CECNL/ExBrainable} for more details.

\section*{Acknowledgements}
This work was supported in part by the Ministry of Science and Technology under Contracts 109-2222-E-009-006-MY3, 110-2221-E-A49-130-MY2, and 110-2314-B-037-061; and in part by the Higher Education Sprout Project of the National Chiao Tung University and Ministry of Education of Taiwan. The authors would like to thank  Xin-Yao Huang for his assistance in parameter optimization.

\bibliography{ExBrainable_arxiv}

\end{document}